\documentclass[12pt,a4paper]{article}

\usepackage{amssymb, amsmath, amsthm}

\usepackage[cm]{fullpage}
\usepackage[english]{babel}
\usepackage[cp1250]{inputenc}
\usepackage[T1]{fontenc}
\usepackage[pdftex]{graphicx}

\title{Derivation of the nonlocal pressure form of the fractional porous medium equation in the hydrological setting}
\author{\L ukasz P\l ociniczak\thanks{Faculty of Pure and Applied Mathematics, Wroc{\l}aw University of Science and Technology, Wyb. Wyspia{\'n}skiego 27, 50-370 Wroc{\l}aw, Poland, Email: \underline{lukasz.plociniczak@pwr.edu.pl}}}
\date{}

\begin{document}
\maketitle

\begin{abstract}
In this short note we consider a nonlinear and spatially nonlocal PDE modelling moisture evolution in a porous medium. We then show that it naturally arises as a description of superdiffusive jump phenomenon occurring in the medium. We provide a deterministic derivation which allows us to naturally incorporate the nonlinear effects. This reasoning shows that in our setting the so-called nonlocal pressure form of the porous medium equation is preferred as a description of the evolution. In that case the governing nonlocal operator is the fractional gradient rather than the fractional Laplacian. \\

\end{abstract}

\section{Introduction}

In \cite{Caf11} the following form of the fractional porous medium equation has been introduced
\begin{equation}
\label{eqn:NonlocalPMECaff}
	u_t = \nabla\cdot\left(u\nabla p\right), \quad p = \mathcal{K}(u),
\end{equation}
where $\mathcal{K}$ is a linear integral operator (for later results see for ex. \cite{Caf13,Sta14,Sta15,Dji18}). Specifically, in this particular case it is the inverse of the fractional Laplacian (the Riesz operator \cite{Ste16,Bil13})
\begin{equation}
\label{eqn:NonlocalPressure}
	\mathcal{K}(u) = \left(-\Delta\right)^{\frac{\alpha}{2}-1}, \quad 0<\alpha<2.
\end{equation}
Not that Authors of the original paper use a different constant $s=1-\alpha/2$. The fractional Laplacian can be defined for example with the Fourier transform $\mathcal{F}$
\begin{equation}
	\left(-\Delta\right)^\frac{\alpha}{2}(v) = \mathcal{F}^{-1}\left(|\xi|^\alpha \mathcal{F}v\right).
\end{equation}
Plugging (\ref{eqn:NonlocalPressure}) into (\ref{eqn:NonlocalPMECaff}) we can see that the following operator arises and can be thought as a fractional gradient
\begin{equation}
	\nabla^{\alpha-1} = \nabla \left(-\Delta\right)^{\frac{\alpha}{2}-1},
\end{equation}
which is a pseudo-differential operator of order $\alpha-1$. It is also possible to define the fractional gradient via the singular integral of a smooth and bounded functions (see \cite{Imb06})
\begin{equation}
	\nabla^{\alpha-1}v(x) = C_{n,\alpha} \int_{\mathbb{R}^n} \left(v(x)-v(x+y)\right)\frac{y}{|y|^{n+\alpha}}dy, \quad x\in\mathbb{R}^n,
\end{equation}
where $C_{n,\alpha}$ is a known constant. In this paper consider the following nonlocal nonlinear diffusion equation
\begin{equation}
\label{eqn:FractionalPME}
	u_t = \nabla\cdot\left(D(u,x)\nabla^{\alpha-1}_xu\right), \quad x\in\mathbb{R}, \quad 0<\alpha<2, 
\end{equation}
Notice that we allow the diffusivity to be a nonlinear function of both the dependent and independent variable. When $D(u,x)\propto u^m$ the above nonlocal PDE reduces to the porous medium equation with a nonlocal pressure.

Apart from the mentioned description of the porous media, generalizations or variations of (\ref{eqn:FractionalPME}) appear is different setting. In \cite{Gia97} a similar nonlocal equation has been used to model long-range interactions in the gas particle system. Moreover, some version of (\ref{eqn:FractionalPME}) has also been used to explain evolution of dislocations in crystal lattice \cite{Hea72}. Furthermore, the so-called hydrodynamic limit appears when $\alpha\rightarrow 0^+$ in which the $1D$ equation (\ref{eqn:NonlocalPMECaff}) can be reduced to the Burgers equation exhibiting hyperbolic shock wave phenomena. Its multidimensional variant has been used for example in modelling vortex liquid in Ginzburg-Landau theory of superconductivity \cite{Wei94}. 

In this short note we give a phenomenological argument that the nonlocal equation of the form (\ref{eqn:FractionalPME}) arises naturally as a description of the moisture imbibition in a porous medium exhibiting superdiffusive jump phenomena. The latter mechanism is responsible for emergence of the spatially nonlocal character of the flux which is represented by the fractional gradient operator. On the other hand, a temporal nonlocality can also arise as a consequence of the waiting time phenomenon in which the water can be trapped in certain regions of the medium for prolonged periods of time. This produces the time-fractional derivative and brings the subdiffusive character of the evolution (for a derivation and related result see \cite{Plo15,Plo14,Plo18}).

In the literature there also exists another form of the fractional porous medium equation, namely
\begin{equation}
\label{eqn:FMPE}
	u_t +(-\Delta)^{\frac{\alpha}{2}}\left(D(u)\right), \quad 0<\alpha<2.
\end{equation}
For a relevant mathematical results see \cite{De10,De12} and for applications see \cite{Bol00,Len03}. Notice that both (\ref{eqn:FMPE}) and (\ref{eqn:FractionalPME}) reduce to the classical porous medium equation in the limit $\alpha\rightarrow 2^-$. On the other hand, in the nonlocal setting, i.e. when $\alpha\in(0,2)$ they also agree in the linear case. Hence, they are nonlocal generalizations of the porous medium equation, albeit it can be shown that they are not equivalent! Most notably, the solutions of (\ref{eqn:FMPE}) have infinite speed of propagation while those of (\ref{eqn:FractionalPME}) are compactly supported. Therefore, the nonlinearity of the equation is the main factor responsible for the lack of unique generalization of the porous medium equation into the nonlocal setting. The interplay of nonlinearity and nonlocality produces a plethora of interesting phenomena. For a detailed comparison and summary see \cite{Vaz14}. In what follows we will show that, at least in the hydrological setting, the nonlocal pressure form is preferred. 

\section{Derivation}

\subsection{Classical case}
Here we revisit the derivation of the classical porous medium equation in the hydrological setting. Let us consider a porous medium (for example soil or brick) and a fluid (water) that penetrates it. We want to derive an equation that governs the evolution of moisture distribution in space and time. By $u=u(x,t)$ denote the fluid concentration (with dimension $M L^{-3}$) at point $x$ and time $t$. Moreover, let a vector $\textbf{q}=\textbf{q}(x,t)$ be the flux (with dimension $M L^{-2} T^{-1}$), that is the amount and direction of the fluid that crosses the unit surface in a unit time. Conservation of mass gives the continuity equation
\begin{equation}
u_t = - \nabla\cdot\textbf{q}.
\label{conti}
\end{equation}
In order to obtain the closure relation we must impose the constitutive equation, which relates the flux to the fluid concentration. It was an experimental fact, proved by Henri Darcy in the middle of $19$th century \cite{Darcy}, that the flux is proportional to the gradient of the pressure (in modern nomenclature)
\begin{equation}
\textbf{q}=-\frac{\rho k}{\mu}\nabla p,
\label{Darcy}
\end{equation}
where $\rho$ is the density of the fluid ($M L^{-3}$), $p$ is the pressure ($M L^{-1} T^{-2}$), $\mu$ is the fluid viscosity ($M L^{-1} T^{-1}$) and $k$ is the geometry dependent permeability ($L^2$). The Darcy's Law is not just an empirical relationship - it can be derived theoretically in a rigorous way by averaging Navier-Stokes equations (which in turn are a consequence of the Boltzmann's kinetic equation, see \cite{Whit86}) In our considerations the fluid density can be assumed to be constant in time and space. This assumption is sensible, since for the pressures and timescales present in our isothermal setting the fluid (water) can be assumed to be incompressible (see \cite{Szym}). The flow is driven by changes in the capillary pressure. This pressure arises due to the surface tension present in the pores. By the Laplace law it can be related to the radius of curvature of the small menisci forming between the surfaces of the solid phase in the medium. This phenomenon is similar to the idealized experiment explaining the surface tension in the capillary (hence the name).

To complete the derivation of the governing equation one must use the mutual dependence of pressure and fluid concentration. For a majority of situations and fluids, there exists a monotone relationship $p=p(u)$, which graph is called the retention curve (for a thorough treatment of the modelling of the flow in the porous media see \cite{Szym}). Using this relationship in (\ref{conti}) and (\ref{Darcy}) we obtain the nonlinear convection-diffusion equation known in hydrology as the Richards equation (see \cite{Rich} and particularly \cite{Bear} for a comprehensive treatment) (in terms of the concentration and without the convection)
\begin{equation}
u_t=\nabla \cdot \left(D(u) \nabla u\right),
\label{Rich}
\end{equation}
where the diffusivity $D$ is defined as $\rho k/\mu \; dp/du$. In almost any case the diffusion in porous media is nonlinear. Moreover, the diffusivity can change over a several orders of magnitude during the imbibition. The choice of the retention relation $p=p(u)$ and hence of the diffusivity $D$ is often taken as to fit the experimental results. One of the most common are the Van Genuhten \cite{vanG}
\begin{equation}
p \propto \left(\frac{1}{u^\frac{1}{m}}-1\right)^\frac{1}{n},
\end{equation}
and the Brooks-Correy functions \cite{BC}
\begin{equation}
p \propto u^{m+1},
\end{equation}
where $n$, $m$ are constants determined from the fitting. In the case of the Brooks-Correy model we obtain the porous medium equation
\begin{equation}
u_t = \nabla \cdot \left(D_0 u^m \nabla u\right),
\label{PMeq}
\end{equation}
where $D_0$ is the diffusion coefficient. The value of $m$ depends on the type of soil investigated and the pore distribution within (typically between $0.2$ and $5$). 

The porous medium equation is also very often derived for a compressible and barotropic gas (see for example \cite{Logan}). For that case one writes the flux as $\textbf{q}=\rho \textbf{v}$, where the velocity is determined from a particular variant of the Darcy's Law. The constitutive equation for the pressure is to assume the thermodynamic equation of state for polytropic process $p\propto u^{m+1}$, which is the same as the Brooks-Correy model in the hydrology. Moreover, a particular form of the porous medium equation arises also in the description of the filtration process, where the water original present in the aquifer, penetrates the porous medium around. The resulting equation for the free-surface is known as the Bussinesq equation (see \cite{Vaz}). In any of these cases we obtain the same equation (\ref{PMeq}) but describing different quantities. 

\subsection{Nonlocal case}
The geometric structure of the porous medium can be very complex. A multitude of different pores, tubes and water filaments of various sizes can introduce many phenomena that can be described by (\ref{PMeq}) only approximately. The description of the transport process in the porous medium has to take into account different (probably all) space and timescales (see \cite{Cush94}) as well as continuously changing heterogeneity scale \cite{Cush93}. Moreover, as is very well-known, the retention curve relation $p=p(u)$ possesses a very strong dependence on the history of the process. Rather than a single curve, the retention relationship is a family of the characteristics different for medium under different moisture conditions. This hysteresis is a clear indicator of the fact that to accurately describe the transport in the porous medium one has to take into account some nonlocal phenomena such as the memory of the process. Lately, a number of experimentalists have shown that the classical diffusion rate cannot accurately predict the moisture evolution in some porous media (like construction materials) \cite{Hal07,Gha04,Loc06}. It was observed that the wetting front moves in a much different pace than the equation (\ref{PMeq}) can predict. This anomalous diffusion is a matter of vigorous research done by both theorists and experimenters. 

As was noted in \cite{Hu94} the classical description of the transport in the porous medium in often inadequate. Moreover, in the same work Authors managed to derive a nonlocal version of the Darcy's law, which is a consequence of the Boltzmann's equation for kinetic transport. It is worth to mention that the nonlocality followed as a necessary condition from the force balance and a very fundamental physical theory. A similar version of the Darcy's Law was presented also in \cite{Mih12}, where the porous medium was modeled with a network of channels of all length-scales through which the water particles can move over a very long distances. The "long jump" property along with the "waiting time" phenomenon, in which the fluid can be trapped for a significant periods of time in some region in space, are the foundations of the anomalous diffusion models that utilize the fractional derivatives. The typical derivation of these models is done in the stochastic framework of the Continuous Time Random Walks (CTRW) \cite{Met00}. An approach based on the conservation of mass was undertaken in \cite{Rina01}. The resulting equation which models the linear nonlocal, anomalous diffusion can we written in the form \cite{Mai07}
\begin{equation}
^C \partial_t^\beta u = -\left(-\Delta_x \right)^\frac{\alpha}{2} u, \quad 0<\alpha<2, \quad 0<\beta<1.
\label{fracDiff}
\end{equation}
Here, the Caputo time derivative accounts for the memory of the process (waiting times) and the fractional laplacian with respect to the space variable models the nonlocal "long jumps" of the fluid particles. In one-dimensional case the fundamental solution of (\ref{fracDiff}) can be expressed in the terms of the Fox H-function. Although the stochastic approach is very illustrative since it utilizes the concept of a randomly walking particle, it is difficult to incorporate the nonlinear diffusion coefficient in it. In what follows we will present a deterministic derivation of the anomalous diffusion equation in the hydrological setting of the porous medium. Our derivation takes into account the nonlinear dependence of the diffusivity on the concentration. 

Following the previous remarks about the necessity of nonlocal phenomena in porous medium we are going to rederive the expression for the flux $\textbf{q}$. For simplicity we will consider the diffusion only in one space dimension. The generalization is straightforward. First, let us discretize time and space by quanta $\Delta x$ and $\Delta t$. Moreover, define the space points by $x_i:= i \Delta x$, where $i\in\mathbb{Z}$. Now, at every $x_i$ consider a three-dimensional box with face area $A$ and height $\Delta x$ and assume that $\Delta x$ is so small, than the concentration of the fluid is almost constant along the box. We ask how much mass goes through the interface from one box to another. If the diffusion were local the flux through the wall with the surface $A$ in time $\Delta t$ would be
\begin{equation}
\textbf{q}=\frac{1}{A}\frac{R(u)}{\Delta t}A \Delta x \left[u\left(x_i-\frac{1}{2}\Delta x,t\right)-u\left(x_i+\frac{1}{2}\Delta x,t\right)\right],
\end{equation}	  
where $R/\Delta t$ ($M T^{-1}$) is the concentration dependent rate of diffusion, that is how much fluid passes the interface in a time $\Delta t$. As a convention, we count the contribution of the fluid parcels flowing from left to right as positive and in the opposite direction as negative. If we assumed that in the limit $\Delta x\rightarrow 0$ and $\Delta t\rightarrow 0$ the quantity $(\Delta x)^2/\Delta t$ would become a constant we would reobtain the classical Fick's Law for the diffusion $\textbf{q}=-D(u) u_x$ and from the continuity equation (\ref{conti}) the governing equation (\ref{Rich}) would follow. Since we assume that the diffusion is nonlocal, the fluid can cross the interface at $x_i$ not only coming from the neighboring boxes but also from any other. In that case we have
\begin{equation}
\textbf{q}=\frac{R(u)}{\Delta t} \Delta x \sum_{j=0}^\infty k_j \left[u\left(x_i-\left(\frac{1}{2}+j\right)\Delta x,t\right)-u\left(x_i+\left(\frac{1}{2}+j\right)\Delta x,t\right)\right],
\end{equation}
where we introduced the weight $k_j$ describing the influence of the box at a distance $j$ from the interface at $x_i$. Notice that we assume that the equally distant boxes contribute the same amount of fluid to the flux at $x_i$. For the local diffusion the coefficients are equal to $k_j=0$ for $j> 0$ and $k_0=1$. We can also write 
\begin{equation}
\textbf{q}=-\frac{R(u)}{\Delta t} \Delta x \sum_{j=-\infty}^\infty \widetilde{k_j} u\left(x_i+\left(\frac{1}{2}+j\right)\Delta x,t\right),
\label{flux2}
\end{equation}
where we introduced the odd coefficients 
\begin{equation}
\widetilde{k_j}\geq 0 \quad \text{for} \quad j\geq 0 \quad \text{and} \quad \widetilde{k_j}\leq0 \quad \text{for} \quad j< 0,
\label{odd}
\end{equation}
while for the classical case we would have $\widetilde{k_0}=-\widetilde{k_{-1}}=1$ and $\widetilde{k_j}=0$ for $j\neq 0,-1$. Now, we take the $\Delta x\rightarrow 0$ and $\Delta t\rightarrow 0$ and formally (\ref{flux2}) becomes
\begin{equation}
\textbf{q}=-D(u,x) \int_{-\infty}^\infty K(x-y)u(y,t)dy,
\label{flux3}
\end{equation}
under the condition that $\Delta x$ and $\Delta t$ goes to $0$ in a way dependent on the sum in (\ref{flux2}). By $K$ we denoted the continuous kernel related to the discreet weight $\widetilde{k_j}$. Equation (\ref{flux3}) is the nonlocal version of the flux derived from assumption that the fluid in the porous medium can contribute to the flux at any point ("`long jumps"') and this contribution depends only on the distance. As the nonlocal flux (\ref{flux3}) has to reduce to the local version in the classical case, we should have
\begin{equation}
\textbf{q}\,_{cl}=-D(u,x) \frac{\partial}{\partial x} u(x,t) = -D(u,x)  \frac{\partial}{\partial x} \int_{-\infty}^\infty \delta(x-y)u(y,t)dy,
\label{cllimit}
\end{equation}
where $\delta$ denotes the Dirac delta distribution. Due to this classical limit we rewrite the nonlocal flux (\ref{flux3}) in the form
\begin{equation}
\textbf{q}=-D(u,x) \frac{\partial}{\partial x}\int_{-\infty}^\infty G(x-y)u(y,t)dy,
\label{flux4}
\end{equation}
where $G$ is some (generalized) function, which particular form we would like to determine. We also note that (\ref{flux4}) is a generalized version of the flux obtained in \cite{Cush91} for a porous medium with an evolving heterogeneity. Since by (\ref{odd}) the kernel $K$ has to be an odd function, $G$ must be even. Moreover, by (\ref{cllimit}) it must approach the Dirac delta function in some limit representing the classical case of local diffusion. One of the simplest choices is to take the power function of the distance, that is $G(y) \propto |y|^{1-\alpha}$ for some $\alpha\in (0,2)$. Specifically, we choose the Riesz potential
\begin{equation}
G(y)=\frac{|y|^{1-\alpha}}{2\Gamma(2-\alpha)\cos\left(\frac{\pi}{2}(2-\alpha)\right)}, \quad 0<\alpha< 2,
\label{Riesz}
\end{equation}
where the constant of proportionality and the parameter $\alpha$ has been chosen appropriately to anticipate further results. It is well-known that so defined $G$ approaches the Dirac delta in a distributional sense when $\alpha\rightarrow 2$, hence $\textbf{q}\rightarrow \textbf{q}\,_{cl}$ as $\alpha\rightarrow 2$. With this representation the nonlocal flux (\ref{flux4}) can be written as
\begin{equation}
\textbf{q}=-D(u,x) \frac{1}{2\Gamma(2-\alpha)\cos\left(\frac{\pi}{2}(2-\alpha)\right)} \frac{\partial}{\partial x} \int_{-\infty}^\infty |x-y|^{1-\alpha} u(y,t)dy=-D(u,x) \nabla^{\alpha-1}_x u(x,t),
\end{equation}
where we have used the definition of the fractional gradient $\nabla^{\alpha-1}_x$ \cite{Bil13,Imb06}. If we had chosen a different prefactor in (\ref{Riesz}) we would obtain a different constant in front of the gradient. This constant could then be easily removed by absorbing it into the diffusivity. Hence, we do not loose any generality by choosing the Riesz potential as a kernel. The resulting nonlocal Richards equation (\ref{Rich}) has the form
\begin{equation}
u_t = \nabla \cdot \left(D(u,x) \nabla^{\alpha-1}_x u\right),
\label{RichNL}
\end{equation}
while the nonlocal generalization of the porous medium equation (\ref{PMeq}) now becomes
\begin{equation}
	u_t = \nabla \cdot \left(D_0 u^m \nabla^{\alpha-1}_x u\right).
\end{equation}
Therefore, at least in the hydrological setting, the nonlocal pressure form of the fractional porous medium equation is well-established from the physical principles. 

\section*{Acknowledgement}
Author would like to express his utmost gratitude to Prof. Grzegorz Karch and Prof. Moritz Kassmann for a lot of prolific talks and discussions. They greatly motivated all the results appearing in this work.

\small

\begin{thebibliography}{10}
	
	\bibitem{Bear}
	Jacob Bear.
	\newblock {\em Dynamics of fluids in porous media}.
	\newblock Courier Corporation, 2013.
	
	\bibitem{Bil13}
	Piotr Biler, Cyril Imbert, and Grzegorz Karch.
	\newblock The nonlocal porous medium equation: Barenblatt profiles and other
	weak solutions.
	\newblock {\em Archive for Rational Mechanics and Analysis}, 215(2):497--529,
	2015.
	
	\bibitem{Bol00}
	Mauro Bologna, Constantino Tsallis, and Paolo Grigolini.
	\newblock Anomalous diffusion associated with nonlinear fractional derivative
	fokker-planck-like equation: Exact time-dependent solutions.
	\newblock {\em Physical Review E}, 62(2):2213, 2000.
	
	\bibitem{Caf11}
	Luis Caffarelli and Juan Vazquez.
	\newblock Nonlinear porous medium flow with fractional potential pressure.
	\newblock {\em Archive for Rational Mechanics \& Analysis}, 202(2), 2011.
	
	\bibitem{Caf13}
	Luis~A Caffarelli, Fernando~Soria de~Diego, and Juan~Luis V{\'a}zquez.
	\newblock Regularity of solutions of the fractional porous medium flow.
	\newblock {\em Journal of the European Mathematical Society}, 15(5):1701--1746,
	2013.
	
	\bibitem{BC}
	AT~Corey.
	\newblock Hydraulic properties of porous media.
	\newblock {\em Colorade State University, Hydraulic Papers}, (3).
	
	\bibitem{Cush91}
	John~H Cushman.
	\newblock On diffusion in fractal porous media.
	\newblock {\em Water resources research}, 27(4):643--644, 1991.
	
	\bibitem{Cush93}
	John~H Cushman and TR~Ginn.
	\newblock Nonlocal dispersion in media with continuously evolving scales of
	heterogeneity.
	\newblock {\em Transport in Porous Media}, 13(1):123--138, 1993.
	
	\bibitem{Cush94}
	John~H Cushman, Xiaolong Hu, and Timothy~R Ginn.
	\newblock Nonequilibrium statistical mechanics of preasymptotic dispersion.
	\newblock {\em Journal of statistical physics}, 75(5-6):859--878, 1994.
	
	\bibitem{Darcy}
	Henry Darcy.
	\newblock {\em Les fontaines publiques de la ville de Dijon: exposition et
		application...}
	\newblock Victor Dalmont, 1856.
	
	\bibitem{De10}
	Arturo de~Pablo, Fernando Quir{\'o}s, Ana Rodr{\'\i}guez, and Juan~Luis
	V{\'a}zquez.
	\newblock A fractional porous medium equation.
	\newblock {\em arXiv preprint arXiv:1001.2383}, 2010.
	
	\bibitem{De12}
	Arturo de~Pablo, Fernando Quir{\'o}s, Ana Rodr{\'\i}guez, and Juan~Luis
	V{\'a}zquez.
	\newblock A general fractional porous medium equation.
	\newblock {\em Communications on Pure and Applied Mathematics},
	65(9):1242--1284, 2012.
	
	\bibitem{Dji18}
	Jean-Daniel Djida, Juan~J Nieto, and Iv{\'a}n Area.
	\newblock Nonlocal time porous medium equation with fractional time derivative.
	\newblock {\em arXiv preprint arXiv:1803.03413}, 2018.
	
	\bibitem{Imb06}
	J{\'e}r{\^o}me Droniou and Cyril Imbert.
	\newblock Fractal first-order partial differential equations.
	\newblock {\em Archive for Rational Mechanics and Analysis}, 182(2):299--331,
	2006.
	
	\bibitem{Gha04}
	Abd El-Ghany El~Abd and Jacek~J Milczarek.
	\newblock Neutron radiography study of water absorption in porous building
	materials: anomalous diffusion analysis.
	\newblock {\em Journal of Physics D: Applied Physics}, 37(16):2305, 2004.
	
	\bibitem{Gia97}
	Giambattista Giacomin and Joel~L Lebowitz.
	\newblock Phase segregation dynamics in particle systems with long range
	interactions. i. macroscopic limits.
	\newblock {\em Journal of statistical Physics}, 87(1-2):37--61, 1997.
	
	\bibitem{Hal07}
	Christopher Hall.
	\newblock Anomalous diffusion in unsaturated flow: Fact or fiction?
	\newblock {\em Cement and Concrete Research}, 37(3):378--385, 2007.
	
	\bibitem{Hea72}
	AK~Head.
	\newblock Dislocation group dynamics iii. similarity solutions of the continuum
	approximation.
	\newblock {\em Philosophical Magazine}, 26(1):65--72, 1972.
	
	\bibitem{Hu94}
	X~Hu and JH~Cushman.
	\newblock Nonequilibrium statistical mechanical derivation of a nonlocal
	darcy's law for unsaturated/saturated flow.
	\newblock {\em Stochastic Hydrology and Hydraulics}, 8(2):109--116, 1994.
	
	\bibitem{Len03}
	EK~Lenzi, RS~Mendes, and C~Tsallis.
	\newblock Crossover in diffusion equation: Anomalous and normal behaviors.
	\newblock {\em Physical Review E}, 67(3):031104, 2003.
	
	\bibitem{Loc06}
	DA~Lockington and JY~Parlange.
	\newblock Anomalous water absorption in porous materials.
	\newblock {\em Journal of Physics D: Applied Physics}, 36(6):760, 2003.
	
	\bibitem{Logan}
	J~David Logan.
	\newblock {\em An introduction to nonlinear partial differential equations},
	volume~89.
	\newblock John Wiley \& Sons, 2008.
	
	\bibitem{Mai07}
	F~Mainardi, G~Pagnini, and Y~Luchko.
	\newblock The fundamental solution of the space-time fractional diffusion
	equation.
	\newblock {\em Fractional Calc. Appl. Anal.}, 4(cond-mat/0702419):153--192,
	2007.
	
	\bibitem{Met00}
	Ralf Metzler and Joseph Klafter.
	\newblock The random walk's guide to anomalous diffusion: a fractional dynamics
	approach.
	\newblock {\em Physics reports}, 339(1):1--77, 2000.
	
	\bibitem{Plo14}
	{\L}ukasz P{\l}ociniczak.
	\newblock Approximation of the erde?lyi--kober operator with application to the
	time-fractional porous medium equation.
	\newblock {\em SIAM Journal on Applied Mathematics}, 74(4):1219--1237, 2014.
	
	\bibitem{Plo15}
	{\L}ukasz P{\l}ociniczak.
	\newblock Analytical studies of a time-fractional porous medium equation.
	derivation, approximation and applications.
	\newblock {\em Communications in Nonlinear Science and Numerical Simulation},
	24(1-3):169--183, 2015.
	
	\bibitem{Plo18}
	{\L}ukasz P{\l}ociniczak and Mateusz {\'S}wita{\l}a.
	\newblock Existence and uniqueness results for a time-fractional nonlinear
	diffusion equation.
	\newblock {\em Journal of Mathematical Analysis and Applications},
	462(2):1425--1434, 2018.
	
	\bibitem{Rich}
	Lorenzo~Adolph Richards.
	\newblock Capillary conduction of liquids through porous mediums.
	\newblock {\em physics}, 1(5):318--333, 1931.
	
	\bibitem{Rina01}
	Rina Schumer, David~A Benson, Mark~M Meerschaert, and Stephen~W Wheatcraft.
	\newblock Eulerian derivation of the fractional advection--dispersion equation.
	\newblock {\em Journal of contaminant hydrology}, 48(1-2):69--88, 2001.
	
	\bibitem{Mih12}
	Mihir Sen and Eduardo Ramos.
	\newblock A spatially non-local model for flow in porous media.
	\newblock {\em Transport in porous media}, 92(1):29--39, 2012.
	
	\bibitem{Sta14}
	Diana Stan, F{\'e}lix del Teso, and Juan~Luis V{\'a}zquez.
	\newblock Finite and infinite speed of propagation for porous medium equations
	with fractional pressure.
	\newblock {\em Comptes Rendus Mathematique}, 352(2):123--128, 2014.
	
	\bibitem{Sta15}
	Diana Stan, F{\'e}lix del Teso, and Juan~Luis V{\'a}zquez.
	\newblock Transformations of self-similar solutions for porous medium equations
	of fractional type.
	\newblock {\em Nonlinear Analysis: Theory, Methods \& Applications},
	119:62--73, 2015.
	
	\bibitem{Ste16}
	Elias~M Stein.
	\newblock {\em Singular integrals and differentiability properties of functions
		(PMS-30)}, volume~30.
	\newblock Princeton university press, 2016.
	
	\bibitem{Szym}
	Adam Szymkiewicz.
	\newblock {\em Modelling water flow in unsaturated porous media: Accounting for
		nonlinear permeability and material heterogeneity}.
	\newblock Springer Science \& Business Media, 2012.
	
	\bibitem{vanG}
	M~Th Van~Genuchten.
	\newblock A closed-form equation for predicting the hydraulic conductivity of
	unsaturated soils 1.
	\newblock {\em Soil science society of America journal}, 44(5):892--898, 1980.
	
	\bibitem{Vaz}
	Juan~Luis V{\'a}zquez.
	\newblock {\em The porous medium equation: mathematical theory}.
	\newblock Oxford University Press, 2007.
	
	\bibitem{Vaz14}
	Juan-Luis V{\'a}zquez.
	\newblock Recent progress in the theory of nonlinear diffusion with fractional
	laplacian operators.
	\newblock {\em Discrete \& Continuous Dynamical Systems-S}, 7(4):857--885,
	2014.
	
	\bibitem{Wei94}
	E~Weinan.
	\newblock Dynamics of vortex liquids in ginzburg-landau theories with
	applications to superconductivity.
	\newblock {\em Physical Review B}, 50(2):1126, 1994.
	
	\bibitem{Whit86}
	Stephen Whitaker.
	\newblock Flow in porous media i: A theoretical derivation of darcy's law.
	\newblock {\em Transport in porous media}, 1(1):3--25, 1986.
	
\end{thebibliography}

\end{document}